# Como o *Game Design* Pode Incentivar o Comportamento Tóxico em Jogos Online

Clara Andrade Pimentel
*Depto. de Comunicação Social*
*Universidade Federal de Minas Gerais*
Belo Horizonte, Brasil
clarapimentel@ufmg.br

Philipe Melo
*Depto. de Ciência da Computação*
*Universidade Federal de Minas Gerais*
Belo Horizonte, Brasil
philipe@dcc.ufmg.br

***Resumo*—Comportamento tóxico é um dos problemas mais associados à comunidade *gamer*. São comuns relatos de discurso de ódio e comportamentos anti-jogo em vários jogos multiplayer online. Muitos estudos abordam essa área, focando nos jogadores e, principalmente, no conteúdo e seu impacto negativo na comunidade. Porém, é possível que outros fatores influenciem a toxicidade. Neste artigo investigamos como elementos do *game design* influenciam o comportamento tóxico, utilizando como base para análise fatores de desinibição tóxica. Para isso, analisamos quatro dos principais títulos do gênero de jogos MOBA com o apoio do *Framework MDA* e de um questionário com jogadores. Os resultados mostram que a maioria dos jogadores tanto já sofreu ofensas como também já ofendeu a outros no jogo, além de reconhecer o MOBA como o gênero que gera mais impaciência entre eles. Além disso, levantamos e discutimos diferentes elementos e mecânicas nos jogos que impactam no bem-estar da comunidade através da individualização dos jogadores, apagamento de solidariedades, e asserção de indiferença a respeito dos comportamentos tóxicos. Com isso esperamos ajudar desenvolvedores e comunidades de jogos a promover dinâmicas de integração.**

## I. INTRODUÇÃO

O comportamento negativo em ambientes online é um dos maiores problemas contemporâneos na web e é um fenômeno muito presente nos jogos digitais online. A toxicidade de comunidades de jogos é amplamente conhecida, tanto por comportamentos individuais hostis, na forma dos *trolls*, quanto por eventos de grande escala, tais como foram o #Gamergate[1] [1], e as mobilizações pela negativação de jogos como The Last of Us II Parte 2 (*NaughtyDog*, 2020) ou Battlefield V (*DICE*, 2018) [2]. Apesar do problema se expandir por toda comunidade *gamer*, é notado que certos gêneros de jogos são mais criticados quando o assunto é comportamento tóxico, com destaque aos jogos do gênero Arena de Batalha Multijogador Online (MOBA).

Num relatório desenvolvido pela Anti-Defamation League e a NewZoo [3], é mostrado que grande parte de milhares de jogadores de 15 jogos online já passaram por experiências de ódio ou assédio. Dota 2 (*Valve Corporation*, 2012) e League of Legends (*Riot Games*, 2009) ocupam lugares de destaque dentre os *games* com maiores taxas de assédio (79% e 75%, respectivamente). O relatório recomenda ações que podem ser tomadas tanto pela indústria de jogos como também sociedade civil, e até governos, para transformar o ambiente *gamer* em algo mais saudável, tais como classificação indicativa, moderação em chat de voz, regras e leis dentro e fora dos jogos, entre outras. Neste sentido, embora a toxicidade seja reportada por jogadores, não é dever exclusivo deles tentar solucionar a questão, sendo ainda uma grande responsabilidade de desenvolvedores e autoridades melhorar a comunidade. Além de agir de forma reativa, através de sistemas de denúncia e métricas de conduta, é necessário avaliar não apenas os *players* e o conteúdo ofensivo, mas lançar um olhar para os processos de desenvolvimento, buscando, com pró-atividade, prevenir o problema e criar jogos com ambientes saudáveis.

As empresas desenvolvedoras dos principais MOBAs já vem implementando diferentes alternativas buscando formas de lidar com o comportamento tóxico. O principal jogo do gênero, League of Legends, possui, por exemplo, o chamado "Código do Invocador", que contém dicas sobre boa convivência. A maioria desses jogos possui também uma ferramenta de denúncias integrada ao jogo para reportar por má conduta e ofensas. Estas ferramentas de denúncia são essenciais para manutenção do ambiente dos jogos online, e podem levar ao silenciamento de contas de jogadores, baixa prioridade em buscas de partida, e até banimento. Não faltam penalizações que podem ser efetivadas. Entretanto, a toxicidade persiste e pode se estender para além do jogo, através da exposição de perfis em redes sociais e perseguições. Embora exista um rol de medidas paliativas que identificam, punem, ou reeducam jogadores com comportamento tóxico, pouco se sabe a respeito das influências dos elementos de jogo, como mecânicas e jogabilidade.

Pensando nisso, este trabalho analisa quatro jogos do gênero MOBA: League of Legends (LoL), Dota 2, Heroes of the Storm (HotS) (*Blizzard Entertainment*, 2015) e *Smite* (*Hi-Rez Studios*, 2014), a fim de investigar como o *game design* pode instigar comportamentos hostis através do próprio jogo e suas mecânicas, e também como podemos estimular uma convivência mais harmoniosa entre jogadores, criando um ambiente propício para tal. Ao invés de focarmos em quem são os jogadores, no conteúdo ofensivo, ou em políticas de punição, nosso objetivo aqui é demonstrar que o comportamento tóxico pode não estar vinculado somente à conduta individual e aspectos culturais ligados ao universo do jogo, mas também com sua própria arquitetura. Embora nenhum aspecto do jogo justifique qualquer tipo de discurso de ódio praticado, identificar os elementos que originam ou permitem o comportamento tóxico é um grande passo no combate contra este tipo de violência online, além colaborar para o desenvolvimento de novos jogos, pensando-os sob as lentes desse problema.

[1]#Gamergate foi uma campanha sistemática de assédio e difamação contra mulheres que trabalham na indústria dos jogos eletrônicos (e.g. desenvolvedoras, jornalistas, jogadoras). Foi popularizado pelo uso de hashtags no Twitter.

Apresentamos neste estudo uma metodologia para análise do *game design* e interface dos jogos sob o viés de como eles impactam na geração de atrito entre jogadores e podem promover comportamentos tidos como tóxicos pela comunidade. Para isso avaliamos comparativamente quatro jogos do mesmo gênero, mas que possuem diferenças marcantes de mecânicas, sob o olhar dos fatores que causam a desinibição tóxica [4]. Isso é feito através da utilização de um consolidado *framework* para estudo e produção de jogos – o Framework MDA [5] – e com o suporte de uma avaliação qualitativa junto à comunidade de jogadores dos jogos analisados, a partir de um questionário online sobre toxicidade. Nossos resultados mostram grande quantidade de jogadores relatando problemas com comportamento tóxico além de evidenciarem mecânicas e dinâmicas que impactam na forma como a toxicidade pode ser performada em cada jogo. Com isso buscamos oferecer apoio para a criação e gerenciamento de jogos, sinalizando possíveis caminhos que minimizem o problema.

## II. Trabalhos Relacionados

Comunidades online vem sendo extensivamente estudadas, seja entendendo como jogadores formam grupos e clãs persistente ao longo do tempo [6]–[8], ou analisando a estrutura funcional de times [9] e dos grupos temporários formados em jogos FPS [10] e MOBAs [11]. Esses trabalhos demonstram que a formação de laços de amizade, ou mesmo contratos temporários de colaboração, estimulam a aprendizagem dos jogadores e produzem bons resultados para os participantes. Em contraste à boa convivência, o comportamento tóxico online também é algo amplamente estudado. Fragoso realiza um estudo sobre o comportamento tóxico do brasileiro em comparação à comunidade internacional com base nas categorias de *spam*, *trollagem* e *griefing* [12], [13]. Em 2005, Warner & Raiter já levantavam questões éticas sobre o abuso de mecânicas e recursos de comunicação em jogos MMO (Multijogador Massivo Online), em especial em *World of Warcraft* (*Blizzard Entertainment*, 2004) e *Toontown Online* (*Disney Interactive*, 2003) [14]. Os recursos nos jogos eram usados para perseguir usuários através de ofensas e atrapalhá-los com habilidades e itens, com foco na prática de *griefing* [15]. Um dos aspectos levantados neste aspecto foi a indiferença dos jogadores em relação aos efeitos reais de suas ações virtuais, e como esse afastamento pode diminuir o senso de responsabilidade sobre seus comportamentos.

Em MOBAs, a relação entre comportamento tóxico e discurso de ódio é ampla, como evidenciado pelas agressões verbais que ocorrem contra jogadores de diferentes raças, etnias, gêneros e religiões [16]. Os estereótipos de gênero perpetuados por personagens em League of Legends ocasionam na perda de confiança de jogadoras e em ofensas sexistas [17], [18]. Por medo de se tornarem alvo, jogadoras podem realizar o auto-apagamento de suas identidades em LoL e Dota 2, ocultando nomes e não utilizando os chats de voz [19]. Os canais de comunicação dos jogos, existentes para a estimular a socialização entre jogadores, muitas vezes, tornam-se um problema para as mulheres, sendo através destes que recebem ofensas e são hostilizadas, o que pode levá-las a deixar de participar das comunidades ou mesmo abandonar o jogo [20]. Sengün et al. [21] também exploram os efeitos negativos do conteúdo de LoL através da análise de inúmeras mensagens de chat com demonstrações de comportamentos tóxicos contra jogadores do Oriente Médio e Norte da África, principalmente alusões ao terrorismo. Outro estudo em larga escala utilizando o Tribunal do LoL [2], envolvendo mais de 10 milhões de denúncias de toxicidade e 1,5 milhões de jogadores, construiu uma base para entendimento do comportamento tóxico no jogo com intuito de auxiliar na detecção, prevenção e medidas para minimizar o problema [22]. Apesar do comportamento tóxico ser amplamente conhecido em LoL e reconhecido como prejudicial para as partidas, grande parte dos jogadores não reconhecem a si mesmos como tóxicos [23], ou seja, não possuem consciência sobre as consequências dos próprios atos.

Embora muitos trabalhos investiguem a toxicidade em jogos, a maioria foca na comunidade, no próprio conteúdo tóxico e nos jogadores, e pouco aprofundam em como o próprio jogo incentiva esse comportamento através de elementos do *game design* que podem provocar tais atitudes. Nessa direção, Fahlström utilizou o framework MDA como metodologia juntamente à etnografia, para realizar uma investigação sobre as mecânicas de League of Legends que promovem efeitos contrários à cooperação de membros de uma mesma equipe em jogo e que produzem frustração em relação aos outros, progredindo até o comportamento tóxico. Como resultado, foram levantados eventos do jogo que desencadeiam toxicidade, ligados aos efeitos negativos de *Competitiveness*, *Individualism* e *High Stakes*, promovidos por mecânicas específicas [24]. De forma similar, em um estudo comparativo entre Counter Strike: Global Offensive (*Valve Corporation*, 2012) e LoL, Al Dafai [25] também utiliza o MDA para mapear elementos similares de design que posicionam ambos os jogos como ambientes de competição profissional. Neste trabalho, foram identificadas afirmações constantes sobre a seriedade da partida e classificação dos jogadores, conferindo a competitividade como fator importante para manutenção do cenário de eSport.

Alguns trabalhos propõem soluções para tentar enfrentar o problema. Silva & Chaimowicz [26] apresentam um agente baseado em inteligência artificial que foi capaz de auxiliar na curva de aprendizado novos jogadores em MOBA tendo em vista a dificuldade que novos jogadores enfrentam devido a hostilidade de comunidades *gamers* online. Ao entender melhor quais elementos do jogo conferem maior atrito entre os jogadores, é possível por exemplo, direcionar esses agentes ou propor mudanças no jogo que melhorariam o ambiente para seus jogadores. Esta pesquisa, portanto, complementa estes estudos anteriores, fazendo uma vasta análise comparativa de toxicidade dentre quatro dos principais jogos MOBA e dando maior foco no entendimento de elementos de mecânica e *game design* que contribuem para o comportamento tóxico dentro do jogo.

[2]Tribunal é um sistema oficial em LoL de avaliação de comportamento tóxico colaborativo, onde os próprios jogadores julgam as situações reportadas.

## III. O Competitivo Cenário dos MOBAS

Multiplayer Online Battle Arena (Arena de Batalha Multijogador Online), ou simplesmente MOBA, é um gênero de jogos eletrônicos que conta com dois times com cinco jogadores cada que se enfrentam até uma equipe conseguir destruir a base do oponente. As partidas são jogadas independentes entre si (como em um jogo de luta). O jogador, em cada partida, se junta com outros 4 jogadores (geralmente desconhecidos) através de um sistema de fila e pareamento[3] determinado pelo próprio jogo. Na partida, ele escolhe um personagem com um conjunto de habilidades e jogabilidade únicas e, junto com o resto do seu time, avança em direção ao time inimigo, tentando destruir suas estruturas de defesa e subjugá-lo, ao mesmo tempo que tenta fortalecer o próprio personagem e defender sua base.

O gênero deriva de um cenário personalizado de jogos de estratégia (RTS), o Defense of the Ancients (DotA), do jogo Warcraft III (*Blizzard Entertainment*, 2002) [27]. Hoje é uma das principais modalidades presentes nos eSports[4] [28], junto a outros como First Person Shoter (FPS) e Battle Royal. Este cenário competitivo conta com jogos consolidados há vários anos: League of Legends (LoL) e Defense of the Ancients 2 (DotA 2), ambos jogados no PC. Outros jogos bem conhecidos são Heroes of The Storm (HotS) e Smite. Juntos eles somam prêmios milionários, movimentando mais de U$ 1 bilhão no ano de 2019[5].

O alto nível de competitividade nos MOBAs gera uma ruptura na forma de jogar, que é integrada ao próprio sistema do jogo: partidas "casuais" e ranqueadas. Embora na prática as partidas sejam equivalentes em termos de mecânica e jogabilidade, a maioria dos jogos possui uma diferenciação no modo de partida que um jogador vai participar. Escolhendo o modo ranqueado, ele enfrentará uma partida em que o resultado irá lhe proporcionar (ou tirar) pontos para um ranking que o classifica em posições dentro do jogo. Essa classificação do jogador dentro do ranking é geralmente usada como medidor de qualidade, servindo muitas vezes como "currículo" para aqueles que almejam um espaço em competições esportivas profissionais.

Jogadores profissionais utilizam o jogo para praticar e treinar suas habilidades, não sendo diferenciados de jogadores "comuns" pelo sistema. Entretanto, mesmo no ambiente profissional, há históricos de casos de pessoas repreendidas por não terem bom comportamento, e várias já foram proibidas de competir após se comportarem de maneira tóxica dentro do jogo. Em 2012, um jogador profissional de LoL foi proibido de competir por um ano após um histórico de abuso verbal no jogo[6]. De forma semelhante, jogadores receberam penalidades semelhantes em 2014 depois de se comportarem de maneira tóxica durante as partidas [7]. Desde então vários outros casos em diversos MOBAs foram se somando aos exemplos de mau comportamento, mostrando a necessidade das empresas em combater este problema.

## IV. Elementos de Toxicidade em Jogos

Utilizamos como orientação para o trabalho o conceito de **desinibição online** [4], fenômeno que ocorre no ciberespaço no momento que um indivíduo se sente desinibido em demonstrar comportamentos que fora do ambiente virtual são considerados divergentes das normas ou mesmo prejudiciais. O fenômeno da desinibição pode proporcionar espaços seguros de interação, onde usuários trabalham seu autoconhecimento e introspecção, mas também pode facilitar a prática do ciberbullying e até mesmo crimes virtuais e discurso de ódio. Suler estabelece 6 fatores que se relacionam para possibilitar este fenômeno: anonimidade dissociativa, invisibilidade, assincronia, introjeção solipsista, imaginação dissociativa, e minimização de status e autoridade. Esses fatores servirão como guia para análise e interpretação dos resultados, buscando assim, elementos no jogo que possam desencadear a desinibição tóxica dos jogadores.

O ambiente online permite a ocultação da identidade offline e a formulação de novas identidades virtuais. Um usuário no ambiente digital é capaz de dissociá-las entre si, constituindo um perfil encapsulado para a rede. Essa fragmentação produz a **anonimidade dissociativa**, a qual pode levar à sensação de invulnerabilidade, pelo fato do usuário poder não ser responsabilizado por suas ações no ciberespaço, ou nem mesmo reconhecer tais ações como parte de sua realidade. A **invisibilidade** dos usuários é a desvinculação do sujeito virtual de uma pessoa real, com ou sem sua identidade conhecida. A predominância do texto e ausência de elementos relacionados à corporeidade e expressão podem gerar efeitos de desumanização e invisibilização da pessoa, e também levar à **introjeção solipsista**, quando cria-se uma representação imaginária e idealizada do interlocutor, atribuindo expectativas à ele que podem ser muito distintas da realidade.

A criação de personagens imaginários e a dissociação da identidade virtual da identidade real fortalecem a **imaginação dissociativa**, que nos jogos é facilmente vista na diferenciação entre o mundo virtual (mundo do jogo) e o mundo real. A **assincronia** diz respeito às diferentes temporalidades que ocorrem no ambiente virtual e o distanciamento do usuário das respostas às suas ações. A ausência de uma resposta imediata aos comportamentos, sensação de invulnerabilidade e a invisibilidade do ciberespaço estão intimamente ligados à

[3]O sistema de pareamento é um sistema automático que encaixa jogadores com níveis e classificações semelhantes no mesmo time.

[4]Termo utilizado para competições organizadas de jogos eletrônicos.

[5]Disponível em https://glo.bo/3fEUH41.

[6]Disponível em: https://lol.gamepedia.com/IWDominate

[7]Disponível em: https://www.gamespot.com/articles/league-of-legends-pro-players-banned-for-toxic-behavior/1100-6420072/

**minimização de status e autoridade**. As pessoas se sentem coibidas a agir quando existe a possibilidade de punição e desaprovação por parte de uma figura de autoridade; se essa figura não existe ou não é visível, há uma predisposição em agir sem restrições.

Fatores culturais e sociais, como a intolerância religiosa e o racismo, são combustíveis para comportamentos tóxicos, como tratado por Sengün et al. [21] em seu trabalho sobre discurso de ódio em LoL expresso através do chat do jogo e em fóruns. Em contraste a isso, alguns jogos conseguem promover o estabelecimento de vínculos de amizade, fortalecendo o trabalho em equipe e incentivando sociabilidades positivas [29], [30]. Considerando os fatores de desinibição tóxica, é possível investigar como o *game design* possibilita agências negativas, e também identificar recursos que fomentam vínculos positivos entre os jogadores, mesmo que temporários, submissos às temporalidades das partidas.

## V. Metodologia

Nossa metodologia consiste em uma análise comparativa de quatro MOBAs (League of Legends, Dota 2, Smite e Heroes of the Storm) através do *framework MDA* (Mechanics, Dynamics e Aesthetics). Buscamos as relações entre o *game design*, mecânicas e toxicidade, explorando a diversidade dos diferentes sistemas dentro de um mesmo gênero de jogo. Para apoiar a análise MDA, contamos também com os fatores de desinibição tóxica [4] e um questionário com jogadores sobre aspectos que a comunidade considera relacionados ao problema.

### A. *Questionário com os Jogadores*

Embora os estudos forneçam direcionamentos sobre toxicidade de algumas comunidades de jogos, ainda é de suma importância a participação dos jogadores durante o processo de entendimento do jogo e suas relações. Portanto, elaboramos um questionário online com o uso da ferramenta Google Forms para jogadores do gênero MOBA considerando os jogos analisados. O questionário consiste em 25 questões de múltipla escolha e uma questão aberta, de sugestões e dúvidas. Assim, buscamos compreender os principais pontos de conflito que emergem dentro do jogo a partir da perspectiva da própria comunidade *gamer*. Através das perguntas, exploramos os sentimentos dos jogadores em relação à toxicidade de suas comunidades, os momentos nos quais é mais comum haver divergências, brigas que resultam em comportamentos hostis e tóxicos, quais os elementos de design envolvidos, e também como os jogadores lidam com este problema. O questionário foi divulgado durante o mês de Julho de 2020 nos principais grupos brasileiros de Facebook relacionados a cada um dos jogos analisados, somando, no final da pesquisa, um total de 313 respostas.

Nessa amostra coletada, observamos primeiro a quantidade de jogadores de cada um dos *games* MOBA apresentados. Embora seja comum um jogador conhecer mais de um desses jogos, é importante saber com qual ele tem mais familiaridade, pois existem diferenças entre cada título que influenciam diretamente nas questões abordadas. A Fig. 1 sumariza quais

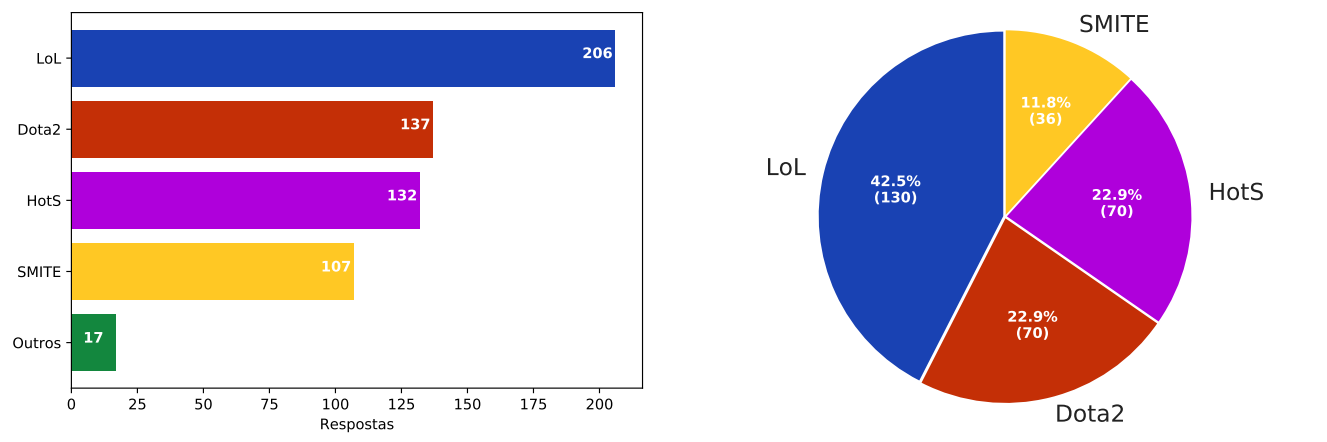

(a) MOBAs que os jogadores jogam ou já jogaram.

(b) Jogo principal do jogador.

Fig. 1. Distribuição dos jogos entre os participantes do questionário em relação aos jogos já jogados e o jogo mais jogado.

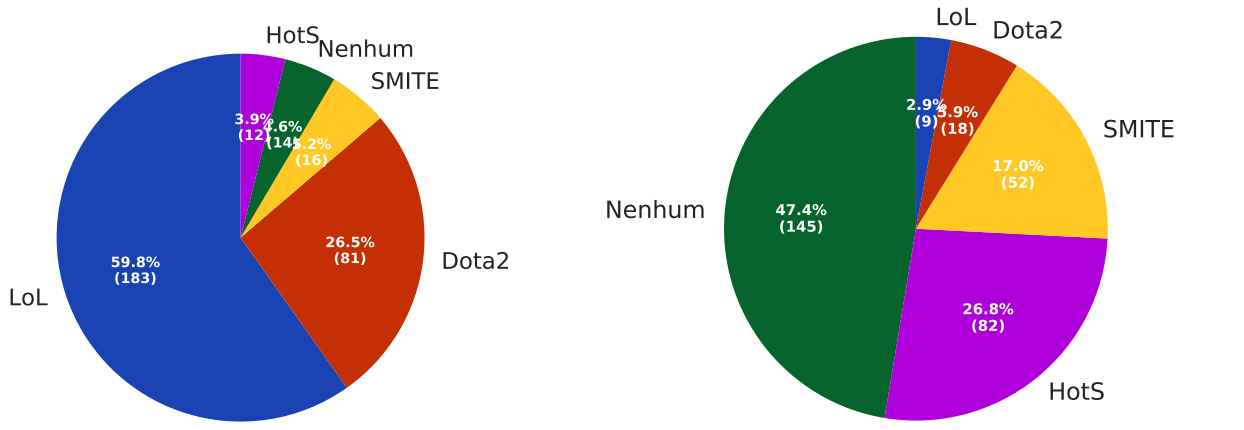

(a) Qual comunidade mais tóxica.

(b) Qual comunidade menos tóxica.

Fig. 2. Comparação de toxicidade entre as comunidades de cada jogo segundo a percepção dos jogadores.

jogos cada jogador já jogou e também o jogo principal de cada um no qual eles possuem maior atividade. Vemos que todos 4 jogos são amplamente conhecidos pelos usuários. A maioria deles tem LoL como seu jogo principal (42,5%), Dota 2 e HotS obtiveram um número semelhante de jogadores (70 - 22,9%) e, por último, 11.8% dos usuários jogam mais Smite. Dentre os quatro jogos, League of Legends se sobressai como o mais conhecido, com 206 pessoas já tendo o jogado ou ainda jogarem.

Ao falarmos sobre toxicidade, perguntamos também a esses jogadores quais desses jogos (LoL, Dota 2, HotS e Smite) eles acham que possuem as comunidades mais e menos tóxicas. A Fig. 2 mostra a visão geral dos jogadores sobre as comunidades. Apenas uma pequena parcela dos jogadores participantes (4.6%) acha que nenhum desses jogos possui comunidade mais tóxica comparado a quase metade (47.4%) que também acredita que nenhuma é menos tóxica, o que mostra que, no geral, esses jogos transmitem uma impressão forte sobre toxicidade. Aqui, é interessante ressaltar que, dentre os jogos, League of Legends foi tanto o jogo mais marcado como "comunidade mais tóxica" como também o menos marcado como o "menos tóxica", por outro lado, de forma inversa, Heroes of the Storm foi o jogo menos marcado como "comunidade mais tóxica" e o mais marcado como o "menos tóxica".

### B. *Framework MDA*

O framework MDA é uma metodologia de análise de jogos digitais baseadas em três categorias: M – **Mecânicas**, D – **Dinâmicas** e A – **Estéticas**. O MDA é uma consolidada estratégia que auxilia desenvolvedores e pesquisadores na análise de jogos digitais, englobando desde os processos de produção, até a experiência do jogador. Nela, o jogo é tratado como um artefato, no qual o "conteúdo do jogo é seu comportamento – não a mídia que é transmitida por ele até o jogador" [5]. Isso nos permite uma compreensão sistemática das potencialidades de interações do ambiente digital e os comportamentos esperados e inesperados resultantes disso. O MDA permite realizar estudos qualitativos sobre efeitos estéticos [31], comparativas entre jogos [25], e até avaliar modelos de gamificação [32]. O método permite sua aplicação a diferentes contextos e sob diferentes demandas, de maneira versátil.

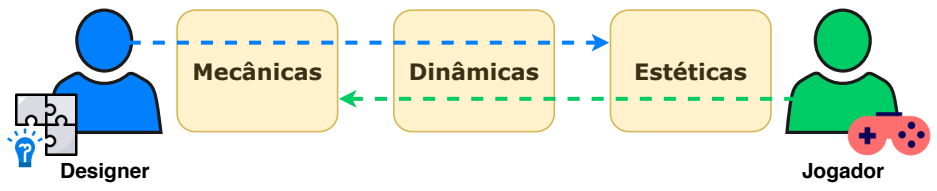

Fig. 3. Diagrama do funcionamento do *Framework* MDA.

As Mecânicas (M) são os diferentes recursos implementados no sistema, em nível de código e informação, que permitem o engajamento do jogador com o programa, tais como interações com objetos, menus clicáveis, ou formas de controle dos personagens. As Dinâmicas (D) são os comportamentos em tempo real resultantes das ações e reações dos jogadores sobre as mecânicas do jogo. Estes comportamentos do sistema e dos usuários desencadeiam uma série de respostas emocionais e experiências, que são englobados pela categoria Estéticas (E). Embora, sejam especificado oito componentes estéticos preliminares, as experiências estéticas do jogador podem abranger diversos campos, inclusive o comportamento tóxico e os fatores de desinibição, utilizados como guia de análise neste trabalho.

Ao aplicarmos o framework fazemos dois percursos distintos, primeiramente partindo da perspectiva do jogador (Estéticas $\rightarrow$ Mecânicas), através da jogabilidade crítica dos MOBAs, observando os comportamentos e suas relações com elementos de design. Para executar essa abordagem, recortamos nossa avaliação ao evento de selecionar e jogar partidas para cada um dos quatro jogos estudados, considerando tanto os momentos de preparação da partida, seu andamento e sua conclusão, além de elementos emergentes paralelos a esse processo, como o sistema de denúncia e o chat durante o jogo. Então traçamos outro caminho na direção Dinâmicas $\rightarrow$ Estéticas, pensando nas agências possíveis dos elementos de jogo e suas consequências. Usando como orientação as formulações de Sueler [4] (descritas na seção IV), investigamos então as estéticas emergentes e suas relações com a desinibição tóxica.

Através desta metodologia buscamos mapear alguns percursos entre toxicidade de partidas, afetos dos jogadores, e as dinâmicas e mecânicas dos jogos. Ou seja, buscamos por dinâmicas que fomentam comportamentos egoísticos e hostis, e ensejam a expressão desses comportamentos.

## VI. Análise de Resultados

A seguir, discutiremos os principais achados a partir de nossa metodologia, ressaltando como os resultados do questionário interagem com os aspectos dos jogos considerados durante a aplicação do framework MDA e relacionados às situações de desconforto, fontes de conflito entre os jogadores, e fatores de desinibição tóxica [4], como a imagem 4 mostra.

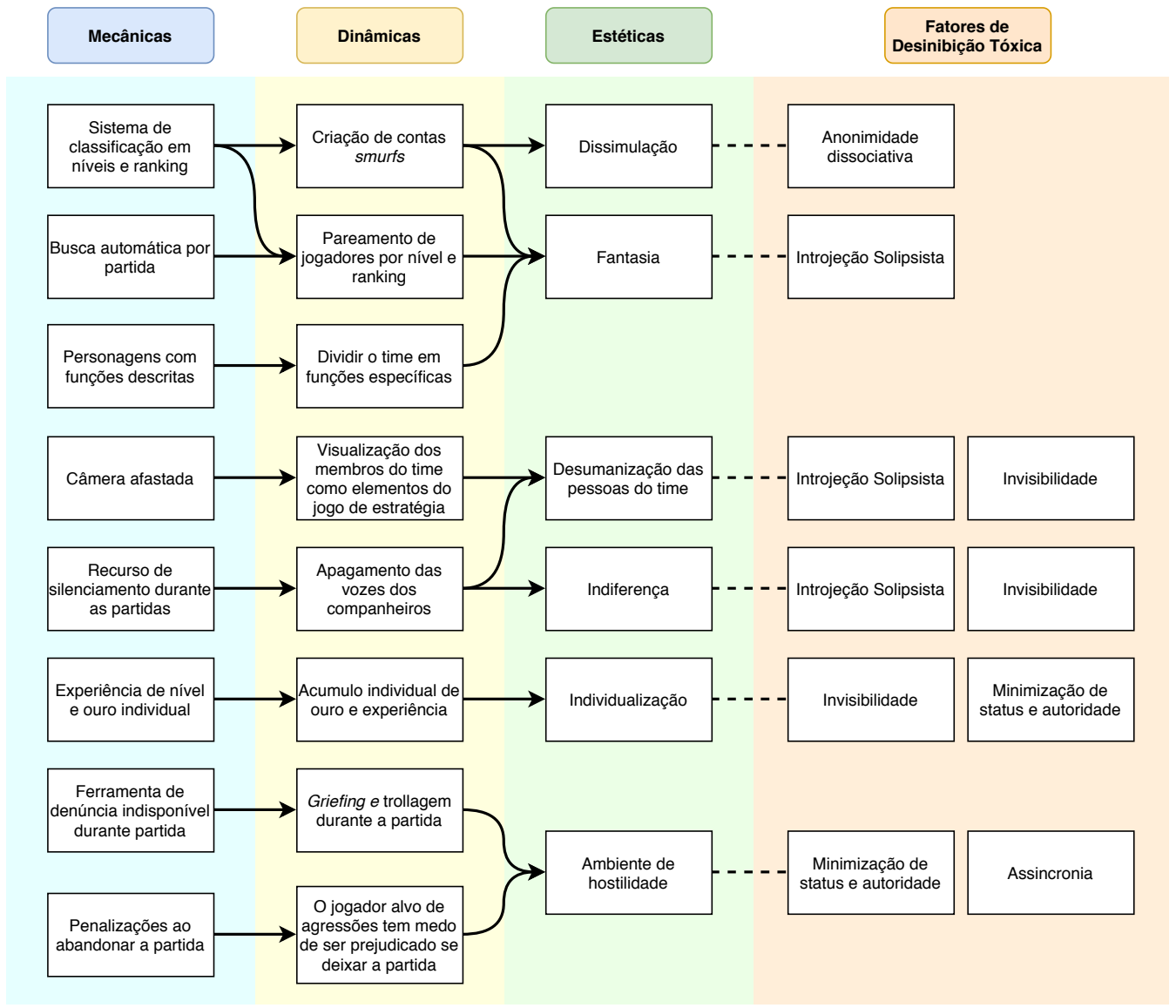

Fig. 4. Diagrama MDA sintetizando os resultados obtidos.

Antes de aprofundar nos elementos do jogo, exploramos os perfis dos jogadores de nossa amostra, exibido na Fig. 5, e suas implicações. Vemos que a maioria dos jogadores (60.5%) se consideram casuais, enquanto 36.9% se veem como jogadores competitivos. Embora os jogos sejam considerados altamente competitivos, inclusive com grandes competições de eSport, observamos através do questionário que a maioria dos jogadores se enxergam como casuais. Essa divisão entre os jogadores é um reflexo da forma como os MOBAs exibem os modos de partida, uma vez que os quatro jogos possuem partidas competitivas/ranqueadas e casuais, o que impacta diretamente no comportamento dos jogadores e em suas expectativas sobre cada partida, mesmo que na prática seja o mesmo jogo.

Outro aspecto do perfil do jogador investigado foi o quanto o gênero MOBA deixa os jogadores mais irritados (Fig. 6). Apenas 32% dos jogadores não ficam impacientes enquanto jogam, enquanto mais da metade dos jogadores (56.2%) afirmaram que o gênero MOBA os deixam mais irritados que outros tipos de jogos, e outros (11.8%) dizem ficar impacientes em qualquer tipo de jogo. Isso mostra que o gênero gera um alto grau de irritação entre os jogadores, mesmo a maior parte deles se considerando apenas jogadores casuais.

O próximo passo na nossa análise MDA foi levantar as mecânicas e elementos de *game design* presentes em cada um dos MOBAs analisados e suas principais diferenças. A Tabela I sumariza os principais pontos levantados nesse quesito. Todos os 4 jogos possuem uma preparação e dinâmica de partida semelhantes: um modo de jogo é selecionado, então, na preparação de partida são escolhidos os personagens e as funções para cada jogador. Iniciada a partida, o time avança no mapa, buscando melhorar seu personagem acumulando os recursos presentes no jogo para sobrepujar personagens inimigos e vencer a partida destruindo a base adversária. Porém, algumas alterações pontuais marcam o ritmo e experiência estéticas em cada um. Entre as diferenças mais marcantes podemos citar a câmera em terceira pessoa em Smite (comparada à câmera isométrica comum de jogos de estratégia nos demais) e a experiência coletiva resultando num nível igual a toda equipe em HotS (comparado ao nível individual por jogador nos outros).

Tabela I
SUMARIZAÇÃO DE SEMELHANÇAS E DIFERENÇAS ENCONTRADAS ENTRE MECÂNICAS NOS MOBAS ANALISADOS.

| **Jogo** | **Divisão de modos de jogo** | **Divisão de funções de personagens** | **Acúmulo de experiência (XP) de personagem** | **Acúmulo de ouro** | **Mecânicas de aprimoramento de personagem** | **Elementos de mobilidade no mapa** | **Comunicação/ Social** | **Tipos de Denúncia** |
|---|---|---|---|---|---|---|---|---|
| **LoL** | Coop. vs IA, PVP, Treinamento, Personalizada | **Funções:** Assassino, Lutador, Mago, Atirador, Suporte, Tanque **Rotas:** Rota Inferior, Rota do Meio, Rota Superior, Suporte, Selva | XP individual de personagem (XP dividida entre todos participantes no abate, priorizando o autor do último golpe) | Ganho passivo, Ouro individual | Nível, Habilidades, Itens | Retorno Automático, Itens | Chat texto e voz, Ping, Emoticons, Dança, Provocação, Piada, Risada | Atitude negativa, Abuso verbal, Abandono de partida/ociosidade, "feeding"intencional, Trapaça, Discurso de ódio, Nome ofensivo |
| **Dota2** | Partidas com Bots, Casual, Competitiva, Salas Personalizadas | **Funções:** Carregador, Suporte, Bombardeador, Desativador, Caçador, Resistentes, Escapista, Empurrador, Iniciadores **Atributos:** Força, Agilidade, Inteligência | XP individual de personagem (XP dividida igualmente entre todos participantes no abate | Ganho passivo, Ouro individual | Nível, Habilidades, Itens, Talentos | Retorno automatico, Itens, Carregador | Chat texto e voz, Ping, Emoticons, Provocação | Abuso de comunicação, Abuso intencional de habilidade, "Feeding" intencional, |
| **HotS** | Contra I.A., Ranqueado, Partida Rápida, Não Ranqueada, Jogos Personalizados | Assassino de Longo Alcance, Assassino Corpo a Corpo, Curandeiro, Suporte, Tanque, Guerreiro | Mesma XP coletiva para todo o time (XP ganhada por participantes em abates) | Sem ouro | Nível, Habilidades, Talentos | Retorno Automático, Montaria | Chat texto e voz, Ping, Emoticons, Sprays, Dança, Falas, Provocação | Conduta abusiva por texto, Conduta abusiva por voz, Morrer de proposito, Ausente/Não participativo, Cheats/Bots/Hacks/Trapaças, Nome impróprio, Spam |
| **Smite** | Normal, Ranqueado, Cooperativo, Treinamento, Personalizado | Guardião, Caçador, Mago, Guerreiro, Assassino | XP individual de personagem (XP dividida igualmente entre todos participantes no abate) | Ganho passivo, Ouro individual | Nível, Habilidades, Itens | Retorno Automático, Itens | Chat texto e voz, VGS (Voice Guided System), Emotes, Pacotes de Voz, Ping | Saindo do jogo, Assédio, Entregando vitória propositalmente, Discurso de ódio, Ameaça real de morte, Não jogou na função atribuida, Outro |

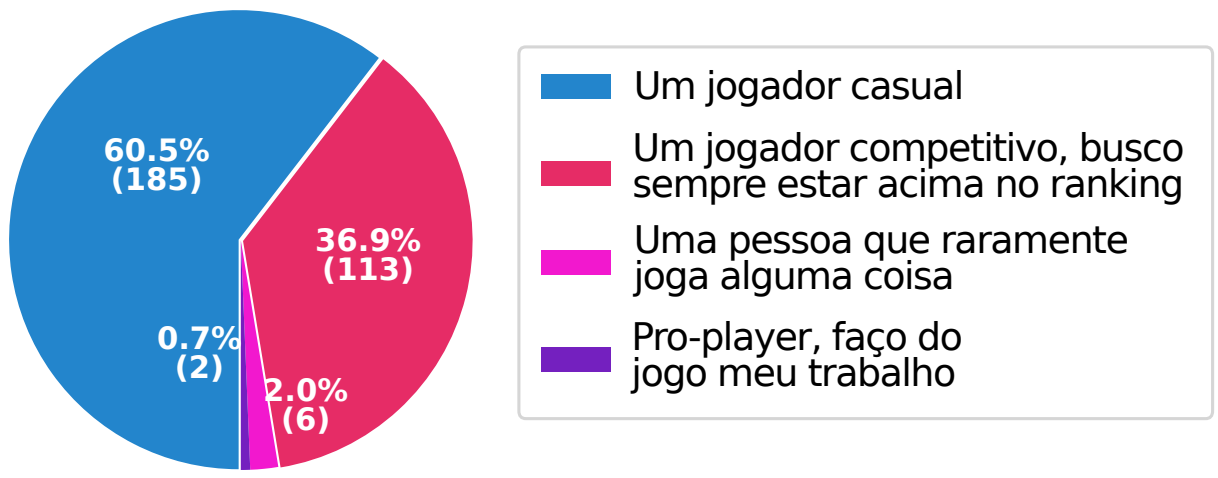

Fig. 5. Dentre as seguintes alternativas acho que me encaixo melhor como.

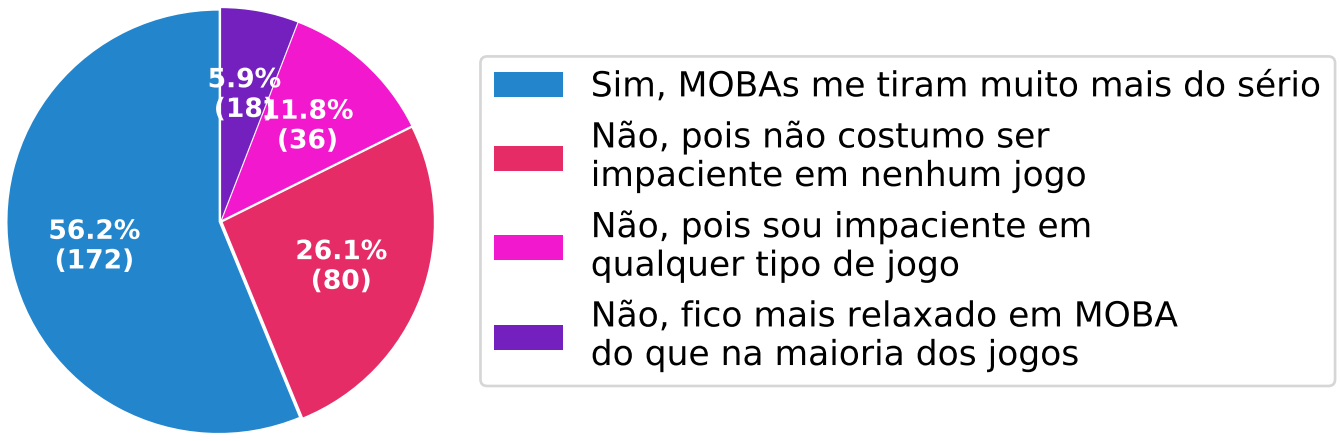

Fig. 6. Você acha que tende a ser mais impaciente quando joga MOBA em comparação a outros jogo?

Com o levantamento das mecânicas, podemos explorar mais a fundo os resultados obtidos pelo questionário e progredir com a análise através do MDA para entender como esses elementos se relacionam com comportamentos tóxicos observados nos jogos. Perguntamos aos jogadores quais são as principais atitudes tóxicas cometidas, baseando-se nos sistemas de denúncias existentes. O resultado nos ajuda a entender como um jogador pode exercer a toxicidade dentro do jogo e como o sistema é capaz de lidar com isso. Como na Fig. 7, o uso do chat para ofensas e o abuso de mecânicas para prejudicar o próprio time são as principais formas de toxicidade apontadas,

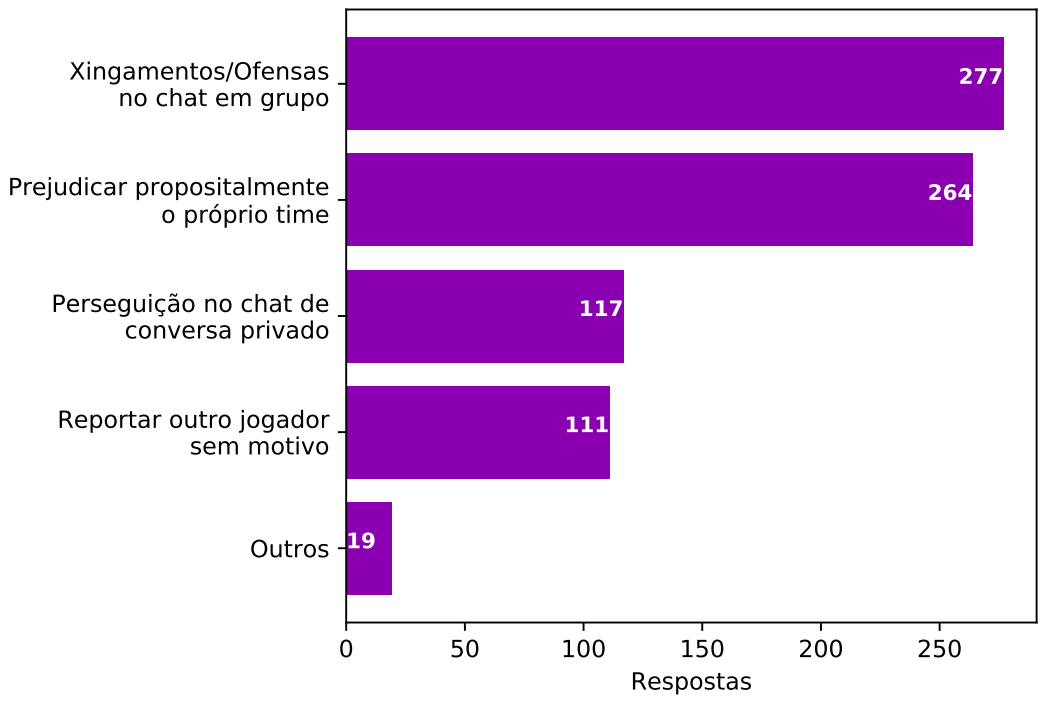

Fig. 7. Quais atitudes de jogadores você acredita que contribuem para a comunidade ser vista como tóxica?

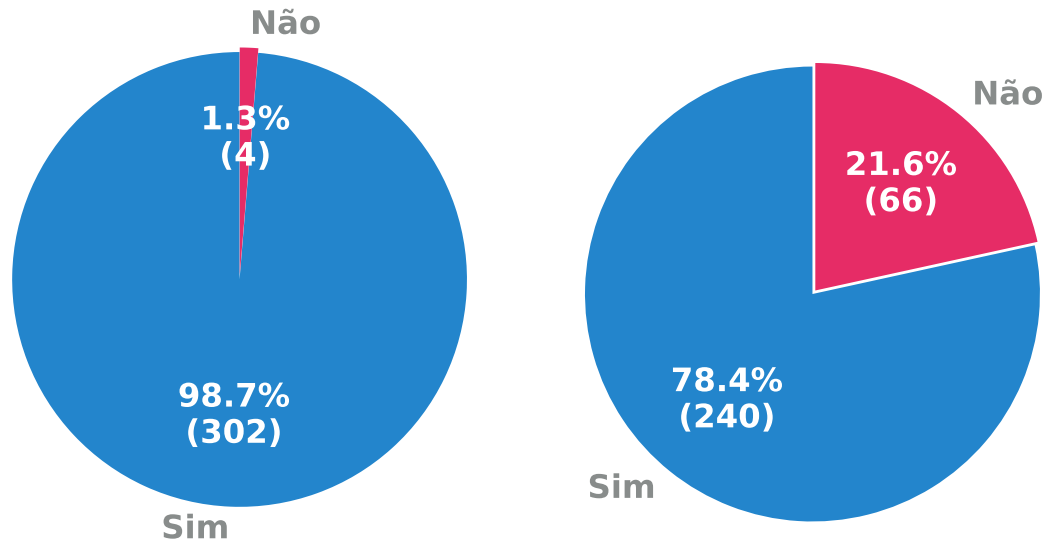

(a) Você já sofreu alguma ofensa dentro do jogo ou em decorrência dele?

(b) Você acredita já ter ofendido outro jogador dentro do jogo ou em decorrência dele?

Fig. 8. Percepção dos jogadores em relação às ofensas dentro do jogo.

porém os jogadores também relatam que o poder de reportar outro jogador pode se tornar um problema quando usado sem um motivo real.

Investigamos também o quanto os jogadores já presenciaram ofensas e outros comportamentos tóxicos dentro do jogo, sendo eles tanto os alvos como também autores de tais atitudes. A Fig. 8 evidencia que ofensas são bem comuns neste ambiente, com quase todos os jogadores (98.7%) já terem vivenciado isso pelo menos uma vez. Mesmo o ato de cometer a ofensa é algo amplamente familiar, uma vez que maioria (78.4%) acredita já ter ofendido outro jogador por causa do jogo. Já na Fig. 9 analisamos o uso por parte dos

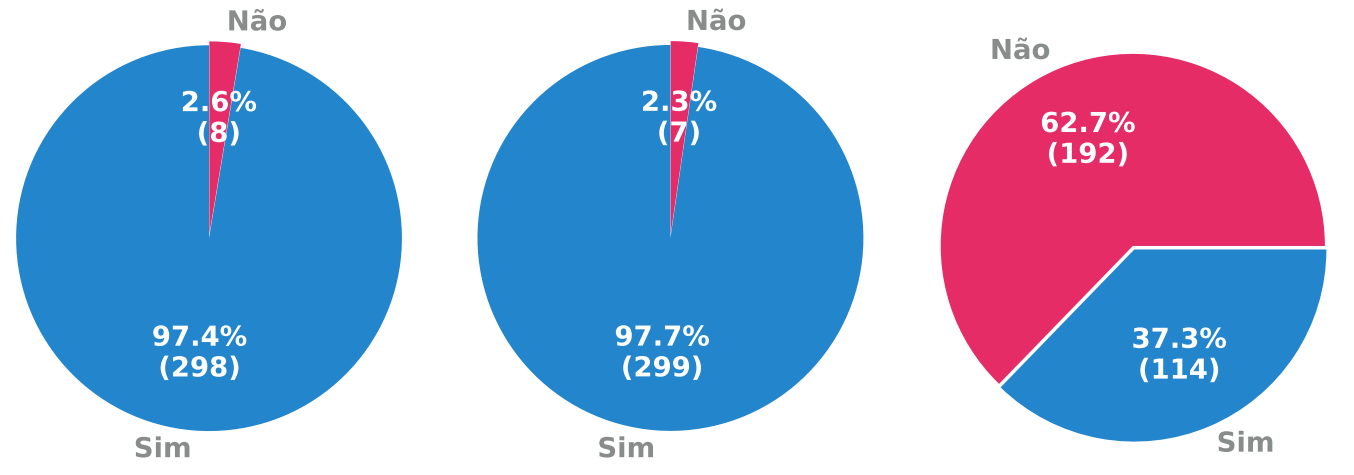

(a) Você já reportou outro jogador por ofensas no chat do jogo?
(b) Você já reportou outro jogador por condutas anti-jogo?
(c) Você já sofreu alguma forma de punição de usuário dentro do jogo?

Fig. 9. Ações efetivas tomadas dentro do jogo em relação a atitudes tóxicas de jogadores.

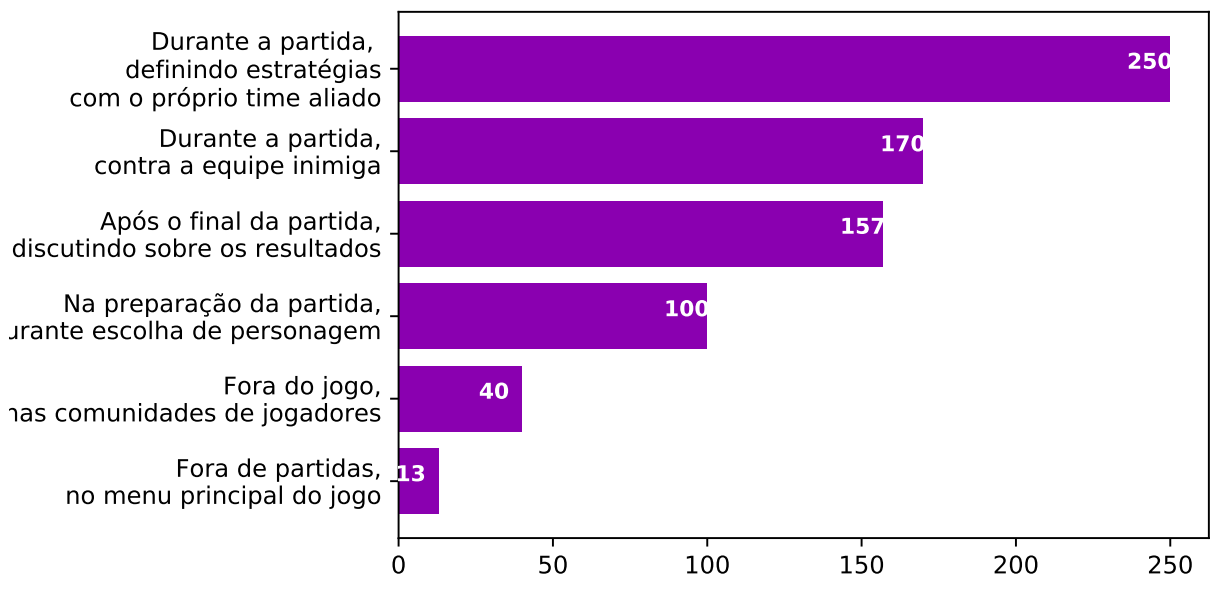

Fig. 10. A maioria dos problemas relacionados aos comportamentos tóxicos acontecem em que momento do jogo?

jogadores das funcionalidades do jogo que lidam com estas condutas tóxicas. Em 9a e 9b, perguntamos aos participantes se já reportaramoutros jogadores e vemos que quase a totalidade ( 97%) já utilizou a ferramenta de denúncia do jogo para reportar alguém, tanto por ofensas no chat quanto por condutas anti-jogo, o que de fato se assemelha a quantidade de usuários que afirmou ter sofrido alguma ofensa. Já em 9c constatamos que mais de um terço dos jogadores já sofreu alguma forma de punição pelo próprio jogo devido má conduta. Isso demonstra que, de fato, uma boa parte dos jogadores também foram reportados em algum momento por atitudes tóxicas, embora as punições sejam em menor escala em relação à percepção de terem cometido alguma ofensa.

Buscamos também entender durante quais momentos do jogo acontecem a maioria dos problemas relacionados aos comportamentos tóxicos. Na Fig. 10, os jogadores apontaram que a toxicidade durante a partida é muito mais direcionada ao time aliado (250 respostas) do que ao time inimigo (170). Com uma quantidade similar de respostas à toxicidade direcionada ao inimigo, há a discussão de resultados após a partida (157). Existe, ainda, um grande número de jogadores afirmando ter problemas relacionados ao comportamento tóxico já na preparação de partida, durante a escolha de personagem (100). Podemos observar ainda que em menor intensidade, a toxicidade se estende até mesmo para fora da partida: no menu do jogo (13) e fora do jogo, em comunidades de jogadores (40). Percebemos com isso que o comportamento tóxico ocorre com maior frequência em relação a colegas do mesmo time.

Pedimos aos jogadores que marcassem de 1 a 5 o quanto alguns eventos relacionados à partida são fonte de atrito no jogo (sendo 1 "não causa atrito nenhum na equipe" e 5 "causa muita irritação na equipe"). Nessas questões, destrinchamos a resposta por jogo para entendermos como as particularidades de cada MOBA impactam na toxicidade, como mostrado nas Figs. 11 e 12. Os resultados nos mostraram que os maiores pontos de atrito compartilhados entre todos são: não levar a sério as partidas ranqueadas (Fig. 11d), perder a partida (Fig. 12d), a presença de um jogador inexperiente (Fig. 11e), e não agir em equipe (Fig. 12e). Esses resultados chamam a atenção para o fato de que apesar da maioria dos jogadores se considerarem casuais, ao somarmos os resultados referentes às partidas ranqueadas, 66% entendem que tais partidas devem ser jogadas com maior compromisso. A percepção sobre jogadores inexperientes coloca em evidência os limites dos sistemas de pareamento, pois nem sempre os níveis de conta e classificações refletem a realidade sobre a habilidade das pessoas.

Algumas respostas apresentam diferenças significativas entre os jogos, mostrando níveis diferentes de atrito que variam de acordo como o jogo trata cada elemento. Um jogador da equipe rival estar mais poderoso causa muito atrito em LoL, enquanto em HotS não (Fig. 12a); matar um inimigo é de extrema relevância em LoL e Smite, enquanto em HotS e Dota 2 não o é (Fig. 12c). Isso é resultado das mecânicas de experiência e ouro nos jogos, que permitem que um jogador da equipe acumule muito mais poder do que os outros. Em LoL e Smite, o jogador que realiza a eliminação do inimigo recebe um bônus de XP e ouro, enquanto os jogadores próximos recebem apenas uma porcentagem da experiência, resultando em uma disparidade entre membros dos times, que possuem níveis diferentes. Em Dota 2 esse desnivelamento é reduzido, devido a experiência por abates e ouro ser igualmente distribuída por todos aqueles que participaram da luta, e ainda menor em HotS, onde não existe nível individual de personagem, e sim para todo o grupo, e não há ouro. Os custos de se perder um jogador com maior nível nesses jogos é bem alto, pois ele se torna um grande pilar de sustentação do time e concede uma recompensa maior quando derrotado, e o abate de inimigos é importante para o fortalecimento de cada personagem.

Por sua vez, a divisão de papéis e posicionamentos na partida gera mais conflito em HotS do que nos outros jogos (Fig. 12b). Isto pode ser explicado pelo fato da divisão de posições no mapa ser menos rigorosa neste jogo em comparação aos outros. Em LoL, por exemplo, cada personagem possui, além de sua função original, indicações sobre em quais rotas deve ficar (e.g. rota superior, rota do meio, rota inferior), que influenciam em como o jogador deve atuar com ele dentro da partida. De forma semelhante, Dota 2 e Smite também possuem uma estratégia muito mais rígida onde cada jogador deve desempenhar um papel específico e pré-determinado antes da partida começar. Em Heroes of the Storm, a presença de vários mapas com diferentes objetivos dificulta o fechamento de uma estratégia única, sendo necessário muitas vezes o time decidir a divisão de tarefas durante o jogo. Embora essa abertura estratégica gere um atrito maior na definição de funções por jogador, por outro lado, papéis muito definidos criam expecta-

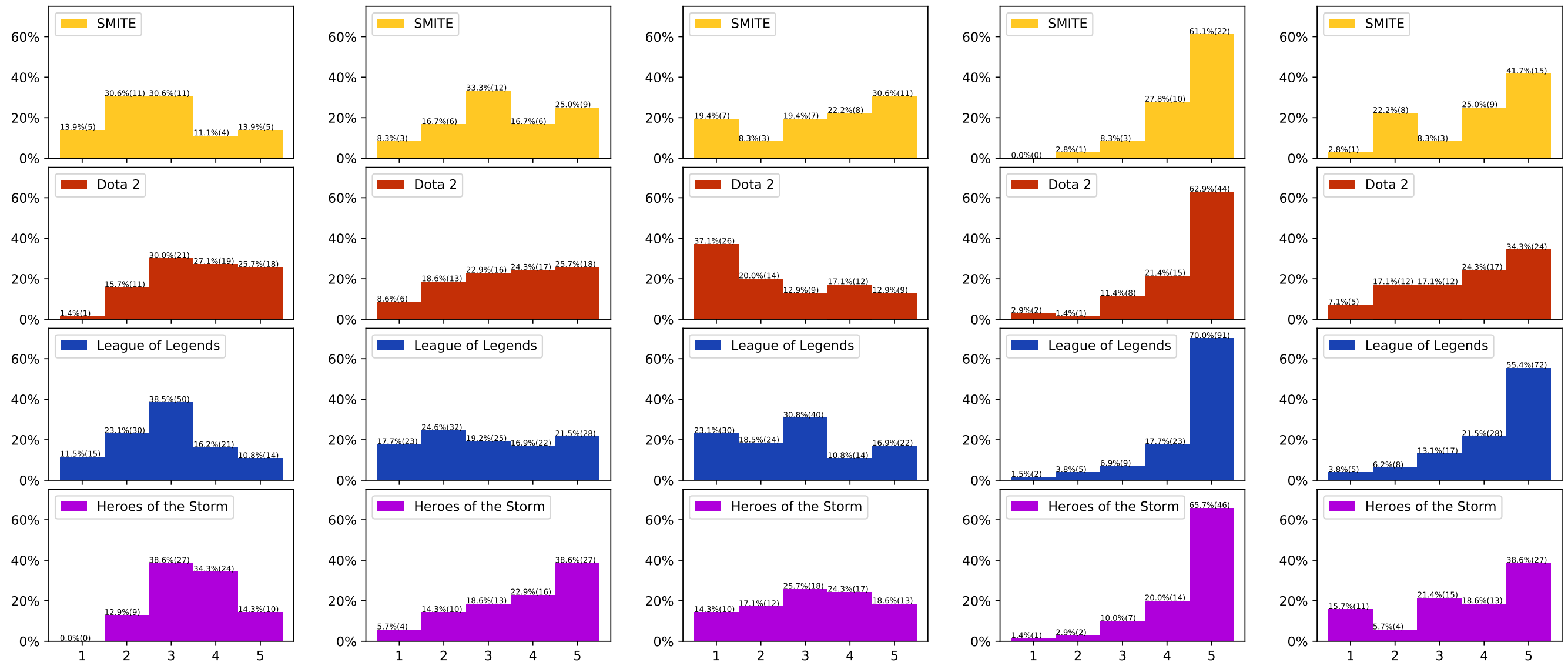

(a) Seleção de heróis para a partida (b) Divisão de papéis e posicionamentos na partida (c) Não levar a sério partidas casuais (d) Não levar a sério partidas ranqueadas (e) Perder a partida tivas sobre como o outro deve se portar. Recordamos aqui um dos fatores da desinibição tóxica, a introjeção solipsista, pois o sujeito cria uma imagem mental e subjetiva do companheiro que conheceu virtualmente, baseada naquilo que deseja que ele faça segundo as estratégias de jogo e interesses pessoais. O resultado disso é uma estética de fantasia baseada nos papéis do jogo, que pode contribuir para a organização do time, mas também ser ponto de atrito quando os jogadores divergem sobre as funções.

Fig. 11. Distribuição de respostas sobre o atrito gerado por diferentes elementos em relação às partidas, separadas por jogo. Valor 1 representa pouco ou nenhum atrito e 5 quando causa muito atrito na equipe.

A dinâmica de parear jogadores por níveis e rankings semelhantes também contribui para a formação de uma fantasia a respeito dos companheiros de time, porque espera-se que tenham habilidades semelhantes às suas. Contudo, o nível ou ranking não traduzem com exatidão como uma pessoa joga. A classificação de usuários ainda produz uma dinâmica de criação de contas múltiplas (as chamadas contas textitsmurfs) para driblar o sistema. As contas nos jogos podem ser criadas com facilidade, usando endereços de e-mail novos, apelidos, e imagens de perfil fictícias. Usuários experientes abusam dessa anonimidade fornecida pelos perfis, e se dissociam das contas antigas por diversos motivos, tais como se livrar das punições acumuladas, ou tentar atingir um ranking melhor mais facilmente. O resultado disso é um ambiente de dissimulação, onde os jogadores podem ser facilmente enganados por outros e suas expectativas sobre as performances dos companheiros não são atendidas. A prática de *smurfing* é considerada um problema por várias empresas de jogos, por atrapalhar o sistema de pareamento e a experiência de jogo dos outros, sendo passível de punição[8].

Uma diferença marcante observada em Smite em relação aos demais é a câmera de jogo, que é mais próxima do personagem neste jogo, enquanto os demais possuem uma câmera isométrica, afastada, semelhante a de jogos de estratégia. A câmera de Smite causa uma sensação de proximidade muito maior ao jogador, trazendo ele ao campo de batalha, mais conectado ao seu personagem e aos dos colegas. Já a afastada produz uma visualização dos personagens como recursos e unidades comuns de jogos estratégicos, invisibilizando que são avatares de jogadores. Esse afastamento causado pela câmera pode levar a uma desumanização dos jogadores, tratados como mero recursos operacionais da partida em vez de pessoas reais batalhando ao seu lado. Outro elemento do jogo que causa afastamento e desumanização entre membros da equipe é o poder de "mutar" os outros jogadores durante a partida. Esse silenciamento, apesar de ser uma das poucas maneiras imediatas de lidar com a toxicidade, permite que todo usuário possa silenciar os companheiros e participar da partida com a ausência de comunicação, reduzindo as pessoas aos seus avatares em tela e ações de jogo. Constatamos que os jogadores que vivenciam toxicidade esperam que outros serão tóxicos e, antecipando isso, jogam com os recursos de comunicação bloqueados, em uma tentativa de afastamento e não afetação. Quem presencia ofensa e toxicidade mas não é alvo pode se afastar da situação de maneira semelhante, optando pela indiferença. Mesmo o chat de texto pode causar essa ideia de distanciamento entre os jogadores, por ser mais desvinculado da pessoa que, por exemplo, o chat de voz.

Como visto, há uma disparidade de poder entre membros de uma mesma equipe, devido à forma como jogadores acumulam XP e ouro. Ao contrário do efeito desejado de camaradagem e trabalho em equipe, isso pode levar a individualização dos integrantes. Apesar haver interdependência entre jogadores

[8]Disponível em: https://glo.bo/2XJVWIW

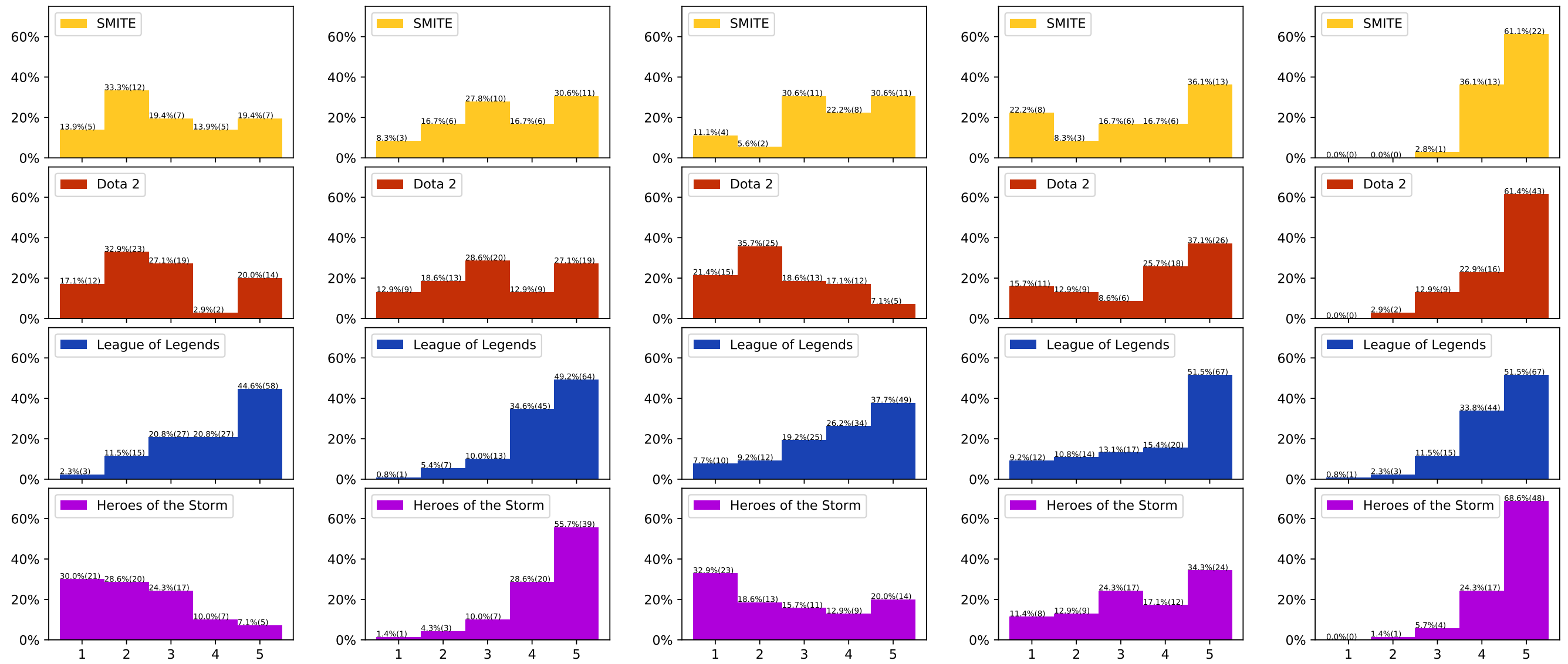

(a) Um jogador da equipe rival estar mais poderoso (b) Perder um objetivo do mapa (c) A *kill* (morte) de um inimigo (d) Um jogador inexperiente com as mecânicas do jogo (e) Um jogador não agir com a equipe e suas funções, é produzida uma ênfase na performance individual. Isso gera um efeito de invisibilidade através da camuflagem da estrutura de grupo, e dificulta a formação de solidariedades entre os jogadores. Além disso, um jogador individual pode adquirir autoridade frente aos demais, através do acúmulo para si de poder ou transferência para rivais através de práticas anti-jogo, como o *feeding* intencional, por exemplo. Esse deslocamento da autoridade diminui as possibilidades de atuação dos integrantes de time, que ficam com a agência submissa ao outro e podem ter seu jogo atrapalhado por outra pessoa.

Fig. 12. Distribuição de respostas sobre quantidade de atrito gerado por diferentes aspectos durante o jogo, separadas por jogo.

Nos jogos, geralmente não há a punição imediata contra o comportamento tóxico durante a partida. O ato de denúncia só traz consequências após a conclusão da partida e depois de ser analisada. Muitas vezes, o próprio autor não sabe nem o resultado que sua denúncia teve, constituindo tempos assíncronos de responsabilização. Por outro lado, a derrota ou mesmo abandonar uma partida no meio, geram penalizações consecutivas para o jogador. Por isso, a autoridade do time pode facilmente ser dissolvida pela ausência de ferramentas que consigam lidar com a toxicidade imediatamente. As penalizações tardias por abandonar a partida fazem com que uma pessoa permaneça no ambiente tóxico, com medo de não poder participar de outros jogos ou perder pontos de classificação. Quem se comporta de maneira tóxica sabe que não pode ser retirado das partidas, e tem liberdade de agir de maneira prejudicial para o time. Isso produz um ambiente de hostilidade em que os jogadores ficam presentes para não receberem punições ou perderem classificações, enquanto outros sabem que podem *trollar* e praticar *griefing*, abusando da assincronia das ferramentas de denúncia e falta de autoridade dos colegas.

## VII. Conclusão

Discurso de ódio e outras violências online são preocupações cada vez mais presentes nas discussões endereçadas não só por pesquisadores, mas também pela própria comunidade *gamer*. Existe uma enorme demanda por soluções para contornar este problema.

Ressaltando os pontos de maior tensão dentre os jogos estudados, essa pesquisa mostra como o *game design* pode estimular o comportamento tóxico nos jogos. Com isso, procuramos auxiliar desenvolvedores, indicando possíveis elementos que podem gerar conflito, mesmo fora do gênero MOBA. Também utilizamos uma metodologia que pode ser replicada em outros jogos para levantar questões relacionadas à toxicidade. Para além do desenvolvimento, conhecendo os pontos de conflito em jogos já produzidos, é possível direcionar medidas de combate ao comportamento tóxico como, por exemplo, criação de conteúdos (vídeos, propagandas, artigos, etc) pela comunidade que possam orientar, ou ainda, com recursos que possam ser implementados nos jogos, semelhante a [26], que cria um agente utilizando IA para auxiliar jogadores novatos.

Neste trabalho comparamos os principais jogos do gênero MOBA na atualidade, analisando a influência dos elementos de mecânica e *game design*. Nossa avaliação qualitativa com os jogadores revelou que a toxicidade é um elemento comum presente nos jogos MOBAs, inclusive gerando mais irritação do que em outros gêneros de jogos. Sobre os elementos motivadores dessa toxicidade, observamos que mesmo sendo *games* do mesmo gênero, diferenças pontuais no design levam a diferentes níveis de toxicidade percebidos pelos jogadores. Nossos resultados mostram que, diferente de muitos jogos cooperativos, há muito atrito entre companheiros de uma mesma equipe, e não em relação à equipe inimiga. Atribuímos isso à alta individualização que pode ocorrer nos MOBAs, que invisibiliza os colegas de equipe e produz um deslocamento da autoridade no jogo. Recursos que promovem a formação de laços de comunidade e o fortalecimento das noções de trabalho em grupo, como guildas/clãs, e opções diferentes de comunicação (*emotes* e animações), podem ser bons investimentos para diminuir a invisibilidade das pessoas nos jogos e promover solidariedades.

Por fim, embora escolhas de *game design* não justifiquem comportamentos tóxicos, principalmente relacionados a preconceitos, como homofobia, racismo ou misoginia, ao entendermos o que permite que jogadores direcionem a outros ofensas ou atrapalhe seus jogos, temos mais recursos para combater a toxicidade e o discurso de ódio, tornando o ambiente mais agradável para todos.

## Referências